\documentclass[aps,amssymb,showkeys]{revtex4}
\usepackage{graphicx}
\usepackage[dvips, bookmarks, colorlinks=true, plainpages = false, citecolor = red, urlcolor = blue, filecolor = blue]{hyperref}
\begin{document}

\title{The density of critical percolation clusters touching
the boundaries of  strips and  squares}

\author{Jacob J. H. Simmons}
\email{Jacob.Simmons@umit.maine.edu}
 \author{Peter Kleban}
\email{kleban@maine.edu}
 \affiliation{LASST and Department of Physics \& Astronomy,
University of Maine, Orono, ME 04469, USA}
\author{Kevin Dahlberg}
\email{dahlberk@umich.edu}
\author{Robert M. Ziff}
\email{rziff@umich.edu}
\affiliation{MCTP and Department of Chemical Engineering, University of Michigan, 
Ann Arbor, MI 48109-2136 USA}

\date{\today}
\begin{abstract}

We consider the density of two-dimensional 
critical percolation clusters, constrained to touch one or both boundaries, in infinite strips, half-infinite strips,  and squares,
as well as several related quantities for the infinite strip. Our theoretical results follow from conformal field theory, and are compared with high-precision numerical simulation. For example, we show that the density of  clusters touching both boundaries of an infinite strip of unit width (i.e.\ crossing clusters) is proportional to 
$(\sin \pi y)^{-5/48}\{[\cos(\pi y/2)]^{1/3}
+[\sin (\pi y/2)]^{1/3}-1\}$.
 We also determine numerically
contours for the density of clusters crossing  squares and long rectangles with open boundaries
on the sides, and compare with theory for the density along an edge.  
\end{abstract}
\keywords{percolation, cluster density, crossing}
\maketitle

\section{Introduction}
Percolation in two-dimensional systems is an area that has a long history, but remains under very active current study. A number of very different methods has been applied to critical 2-D percolation, including  conformal field theory (CFT)  \cite{Cardy92}, other field-theoretic methods \cite{BD}, modular forms \cite{KlebanZagier03}, computer simulation \cite{KZi}, Stochastic Lo\"wner Evolution (SLE) processes \cite{JD06} and other rigorous methods \cite{MA}.  (Because the literature is so very extensive we have cited only a few representative works.)  

More specifically, there is a  great deal of work of recent work
studying universal properties of
crossing problems in critical percolation in two dimensions
(i.e., \cite{LanglandsEtAl92,Cardy92,S01,BercheDebierreEckle94,Watts96,JD06,Ziff92, KlebanZagier03,Vasilyev06}).
Another interesting and also practically important universal feature of
percolation at the critical point is the density,
defined as  the number of times a site belongs to clusters satisfying some specified boundary condition (such as clusters touching certain parts of the boundary)
divided by the total number of samples $N$,
in the limit that $N$ goes to infinity. 
This problem has been addressed   for clusters touching any part of the boundary of a system in various
geometries, including rectangles, strips, and disks, via conformal field theory \cite{BE85} and by solving the problem for
a Gaussian free field and then transforming to other statistical mechanical models, including percolation
 \cite{ResStraley00}.  
  In a recent Letter \cite{KlebanSimmonsZiff06}, we
considered the problem of clusters simultaneously touching one or 
two intervals on the boundary of a system, and considered cases where
those intervals shrink to points (anchor points).  These results 
exhibit interesting factorization, are related to two-dimensional  electrostatics, 
and highlight the universality of percolation densities.  Note that the density at a point $z$ of clusters which touch specified parts of the boundary is proportional to the probability of finding a cluster that connects those parts of the boundary with a small region around the point $z$.

In this paper we consider the problem of the density $\rho_b$ of critical percolation
clusters in various geometries where the clusters simultaneously touch
{\it both} of the boundaries (i.e. crossing clusters), and several related quantities. 

The first case we consider is an infinite strip, with boundaries parallel to the $x$-axis at
$y = 0$ and $y = 1$, so the crossing is in the vertical direction.  (All our models are defined  so that the crossing is vertical. Fig.\ \ref{figx} below illustrates the geometries that we consider.)  For the infinite strip we find (leaving out an arbitrary normalization constant
here and elsewhere) that 
\begin{equation}
\rho_b(y)=(\sin \pi y)^{-5/48}\left[\left(\cos \frac{\pi y}{2} \right)^{1/3}
+\left(\sin \frac{\pi y}{2} \right)^{1/3}-1\right] \ .
\label{rhoB}
\end{equation}
This may be compared to a previous result \cite{BE85,ResStraley00} for clusters touching {\it either the upper or lower boundary (or both)} 
which is simply given by
\begin{equation}
\rho_{e}(y)=(\sin \pi y)^{-5/48} \label{rhoe} \ .
\end{equation}
We also show that the density of clusters touching
{\it one boundary irrespective of touching the other}
is given by
\begin{eqnarray}
\rho_0(y)&=&(\sin \pi y)^{-5/48} \left( \cos \frac{\pi y}{2} \right)^{1/3} \label{rho0} \\
\rho_1(y)&=&(\sin \pi y)^{-5/48} \left( \sin \frac{\pi y}{2} \right)^{1/3} \label{rho1y} \ ,
\end{eqnarray}
where $\rho_0$ corresponds to those clusters touching
the lower boundary at $y = 0$, and $\rho_1$ corresponds to those
clusters touching the upper boundary at $y = 1$.  (Note that $\rho_0$ is the analog of the order parameter profile $\langle \sigma \rangle_{+,f}$ for the Ising case (see Eq. (16) in \cite{CFT91a}.)
We also find expressions for clusters touching one boundary
and not the other, which are combinations of the above results.

Perhaps the main new theoretical result in the above is (\ref{rho0}) (or equivalently  (\ref{rho1y})), which follows straightforwardly from the results in \cite{KlebanSimmonsZiff06}.  The derivation is given in section \ref{infth}.

A second type of theoretical prediction gives the density variation along a boundary (the general expressions is in (\ref{edgetheory})).  This is used to predict the density along the edge in several geometries (see  (\ref{edgetheory2}), (\ref{rhoBygg0}), and (\ref{rhoBx0}) below).

The above theoretical results are found to be consistent with numerical simulations to a high degree of accuracy.  We include the results of numerical simulations of the density contours of clusters crossing square and rectangular systems vertically, with open boundaries on the sides, and compare  with  theory along the boundaries.

Our theoretical treatment is related to previous use of conformal field theory to study order parameter profiles  in various 2-D critical models with edges and similar research \cite{BE85, CFT87a, CFT87b, CFT87c, CFT90, CFT91a, CFT91b, CFT94, CFT97a, CFT97b, CFT98, CFT00a,  CFT00b, CFT00c, 
CFT01}. (Note that the density which we consider is the expectation value of the spin operator in percolation, which is the order parameter in this setting.)   Many of these prior results make use of  the original Fisher-de Gennes proposal \cite{FdG} for the behavior of the order parameter at criticality near a straight edge.  In this paper, we limit ourselves to critical percolation. We also include the results of high-precision computer simulations.  In addition, the formula (\ref{edgetheory}) for the density along the edge of a system is new, to our knowledge.

Note that, 
because of the fractal nature of critical percolation clusters,
the density of clusters is,  strictly speaking, zero everywhere in the system.
However, if we properly renormalize the density as the
lattice mesh size goes to zero, the density can remain finite. At the boundaries, for some quantities,
this results in the density diverging, as for $\rho_e(y)$ at $y = 0$
and $y = L$  (but remaining integrable).
 For  $\rho_b(y)$ of equation
(\ref{rhoB}), on the other hand, the renormalized
density remains finite everywhere.
 When comparing to numerical
simulations, one has to normalize the data so that the densities
coincide with the theoretical prediction using whatever normalization
 convention is chosen for the theoretical results.  The resulting normalization constant, which must be applied to
the numerical data, is specific for each system and is non-universal.

In the following sections, we first give the theoretical derivation of our
 infinite strip formulas above.  Then we present the numerical results. This is followed by numerical results on square and (long) rectangular systems.  These are compared with theory for the density along the edges of these systems.  We end with a few concluding remarks.

\section{Theory for the infinite strip} \label{infth}

We first consider the density of critical percolation clusters which span the sides of an infinite 2-D strip.  We can find the density predicted by conformal field theory \cite{BPZ84} using the results of
 \cite{KlebanSimmonsZiff06}.
 In that article we showed that in the upper half plane the density $\rho$ of clusters connected to an interval $(x_a, x_b)$ is 
\begin{equation}\label{rho1}
\rho(z,x_a,x_b) \propto (z-\bar{z})^{-5/48}F\bigg(\frac{(x_b-x_a)(\bar{z}-z)}{(\bar{z}-x_a)(x_b-z)}\bigg) \ ,
\end{equation}
where the function $F(\eta)$ was determined by conformal field theory and takes on one of two forms,
\begin{equation}\label{rho1a}
F_{\pm}(\eta) = \bigg(\frac{2-\eta}{2\sqrt{1-\eta}}\pm 1\bigg)^{1/6} \ .
\end{equation}

If we parameterize $z$ as $z=r e^{i \theta}$ and let $x_a \to 0$ and $x_b \to \infty$, then $\eta=1-e^{2 i \theta}$ and using (\ref{rho1a}) we can rewrite (\ref{rho1}) as
\begin{eqnarray}
\rho_+(r,\theta,x_a\to0,x_b\to\infty)&\propto&(r \sin\theta)^{-5/48} [\cos(\theta/2)]^{1/3}\\
\rho_-(r,\theta,x_a\to0,x_b\to\infty)&\propto&(r \sin\theta)^{-5/48} [\sin(\theta/2)]^{1/3} \ .
\end{eqnarray}

For the positive real axis $\theta \to 0$ and $\rho_+ \sim \theta ^{-5/48}$ while for the negative real axis $\theta \to \pi$ and $\rho_+ \sim (\pi-\theta)^{11/48}$.  The powers here arise from the fixed and free boundary exponents, respectively, in the bulk-boundary operator product expansion of the magnetization operator $\psi$ \cite{KlebanSimmonsZiff06}. (More specifically, as it approaches the boundary, $\psi \sim \mathbf{1}$ or $\psi \sim \phi_{1,3}$, which have conformal dimensions $0$ and $1/3$, respectively.) This shows that $\rho_+$ is the density of clusters attached to the positive axis.  Because $\rho_-(r,\theta)=\rho_+(r,\pi-\theta)$, it follows that $\rho_-$ is the density of clusters attached to the negative real axis.

The final density that we need is that of clusters connected arbitrarily to the axis.  This is given by $\langle \psi(z,\bar{z}) \rangle_{\rm fixed} \propto (z-\bar{z})^{-5/48}$
\cite{KlebanSimmonsZiff06,BE85,ResStraley00} which may also be written
\begin{equation}
\rho_a(r,\theta) \propto (r \sin\theta)^{-5/48} \ .
\end{equation}

These densities are unnormalized. However  for points that are short distances above the positive (negative) axis, but very far from the origin, the relation $\rho_{+(-)} \approx \rho_a$ holds.  This condition holds since the points are far from the free boundary, and thus dominated by the fixed boundary.  It is satisfied by our expressions for $\rho_+, \rho_-,$ and $\rho_a$, so  they are properly normalized relative to one another.

We next map these densities into the infinite strip 
$w \in $\mbox{$\{x+i y|$} \mbox{$x \in (-\infty,\infty)$}, \mbox{$y\in(0,1)\}$} using
\begin{equation}
w(z)=\frac{1}{\pi}\log(z) \ .
\end{equation}
This leads to the expressions for $\rho_0(y)$, $\rho_1(y)$, and $\rho_e(y)$ given by 
equations (\ref{rho0}), (\ref{rho1y}), and (\ref{rhoe}), respectively.

Using these functions we can also determine the densities of clusters that touch {\it one side but not the other},
\begin{eqnarray}
\rho_{0\bar{1}}(y)&=&\rho_e(y)-\rho_1(y)=(\sin \pi y)^{-5/48}\left(1-[\cos(\pi y/2)]^{1/3}\right) \label{rho01}\\
\rho_{1\bar{0}}(y)&=&\rho_e(y)-\rho_0(y)=(\sin \pi y)^{-5/48}\left(1-[\sin(\pi y/2)]^{1/3}\right) \label{rho10} \ .
\end{eqnarray}

In a similar manner we can find the density of clusters touching both sides,  $\rho_b(y)$. Adding $\rho_0$ and $\rho_1$ includes all configurations that touch either side, but double counts the clusters that touch both sides.  Subtracting $\rho_e$ leaves only those clusters that  touch both sides of the strip.  Thus
\begin{equation}
\rho_b(y)=\rho_0(y)+\rho_1(y)-\rho_e(y)\;,
\label{rhob}
\end{equation}
which gives  equation (\ref{rhoB}).

\begin{figure}
\begin{center}
\includegraphics[width=5in]{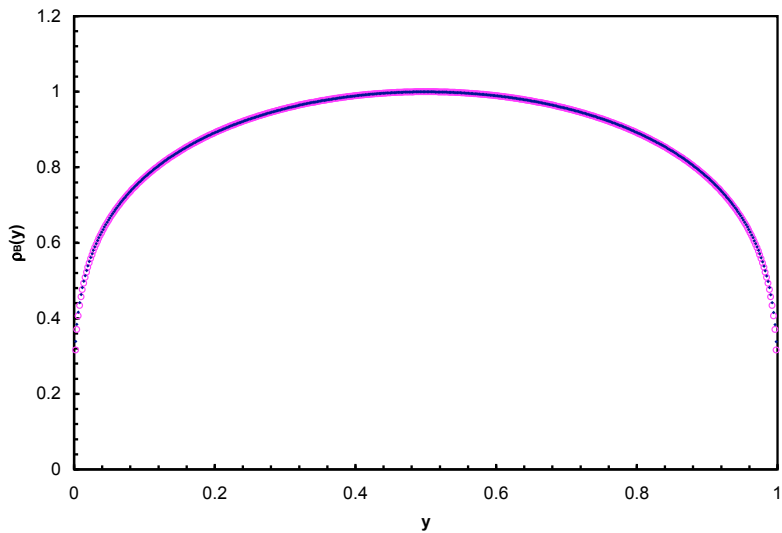}
\caption{(Color online).} $\rho_b(y)$ vs.\ $y$. 
Open violet circles,
theoretical values from equation (\ref{rhob}) (here normalized to $1$ at the center point), using (\ref{yformula})
with $a = 0$.
Blue dots:  results from simulations. \label{figrhoBa0}
\end{center}
\end{figure}

\begin{figure}
\begin{center}
\includegraphics[width=5in]{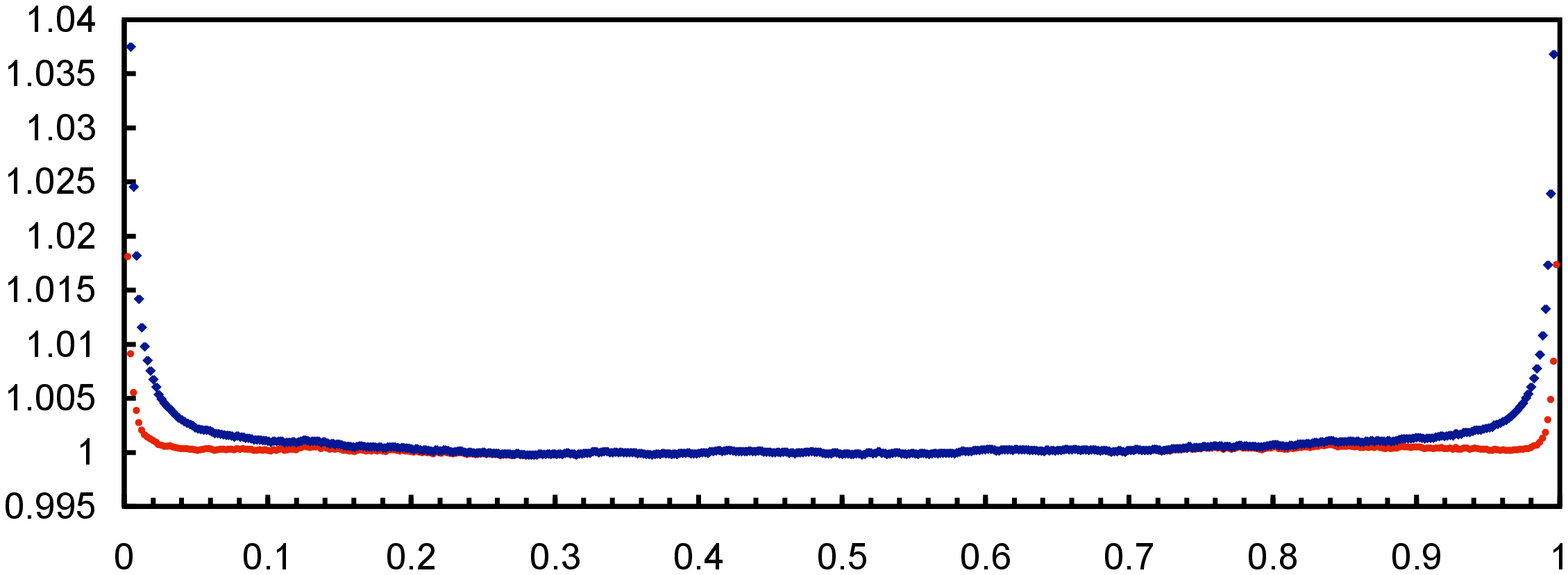}
\caption{(Color online). Ratio of simulation to theoretical
results for $\rho_b(y)$ with $a = 0$ (upper set of points)
and $a = 0.26$ (lower
set of points)}.\label{figratioboth}
\end{center}
\end{figure}

\section{Simulations for the infinite strip}
To approximate the infinite strip, we considered rectangular systems
with periodic boundary conditions in the horizontal direction, for both
site and bond percolation on square lattices.  Here we report our results
for site percolation on the square lattice, for a system of 511 (vertical)
$\times$ 2048 (horizontal) sites,
at $p = p_c = 0.5927462$ \cite{NewmanZiff00}.  We generated 500,000 samples
to compute the average densities, using a Leath-type of algorithm to find all clusters
touching the upper and lower boundaries.

 The aspect ratio of the rectangle used in our simulations was $2048/511 = 4.008 \ldots$, which might seem a bit small.  However, since the correlation length of the system is governed by the width of the rectangle ($511$), the effect of the finite ratio drops off exponentially with the length of the rectangle, so our results  should be very close to those of an infinite strip, an expectation which is born out by the results described below. (Furthermore,  the probability of finding a horizontally wrapping cluster drops exponentially with the aspect ratio, so if a longer rectangle were used, very few wrapping clusters would be found and the data would have poor statistics.)

The agreement of predicted and simulated values is excellent.  In Fig.\ \ref{figrhoBa0}
we show the results for $\rho_b(y)$ with no adjustments, other than an overall normalization (in particular, the extrapolation length $a$, described below, is set to zero).  Fig.\ \ref{figratioboth} (upper set of points) shows the ratio of simulation to 
theoretical results, equation (\ref{rhob}).  Most points fall within 
1\%, except those right near the boundary.  However, we can do better.

When comparing simulations on a (necessarily) finite lattice with the results of a continuum theory, there is the question, due to finite-size and edge effects, of what value of
the continuous variable $y$ should correspond to  the lattice
variable $Y$ --- specifically,
where the boundary of the system should be placed.  
On the lattice, the density goes to zero at  $Y = 0$ and $Y = 512$. 
A na\"ive assignment of the continuum coordinate would therefore
be  $y = Y/512$.  However, we can get a better fit to the data near the
boundaries by assuming that the effective boundary is a distance $a$
(in units of the lattice spacing) 
beyond the lattice boundaries --- that is, at  $Y = - a$ and $Y = 512 + a$.
The distance $a$ 
is an effective ``extrapolation length" where the continuum 
density far from the wall extrapolates to the value zero \cite{Ziff96}.  
This is accomplished by defining
the continuum variable $y$ by 
\begin{equation}
y = (Y + a)/(512 + 2a) \ .
\label{yformula}
\end{equation}
Then,  $Y = -a $ corresponds to $y = 0$ and $Y = 512 + a$
corresponds to $y =1$.   Note, $Y = 0$
corresponds to $y = a/(512 + 2a)$ and $Y = 512$ corresponds to 
$y = (512 + a)/(512 + 2a) = 1 - a/(512 + 2a)$, and  so the 
theoretical extrapolated density $\rho_b(y)$ will
be greater than zero at these points  on the actual boundary.  The spacing between
all points is stretched by a small amount because of the
denominator in equation (\ref{yformula}), but this stretching does not have
much effect on the behavior of $\rho_b$ near the center.  The main effect of $a$
on the shape of the theoretical curves of $\rho_b$ is near the boundaries.

By choosing an extrapolation length of $a = 0.26$, we can get a much
better fit of the data, as can be seen in Fig.\ \ref{figratioboth}
(lower set of points).  A plot of the data analogous to Fig.\ \ref{figrhoBa0}
puts most of the data  points right in the center of the theoretical
circles, but would barely be visible  when plotted on this scale.
With $a = 0.26$, the error is now reduced to less than 0.1\%,
except right near the boundaries, as can be seen
in Fig.\ \ref{figratioboth}.    A more thorough study of
the extrapolation coefficient $a$ would require the study
of different sized lattices, and the demonstration that $a$
is independent of the lattice size.  We have not carried 
this out. Note, however, that a constant $a$ implies that if one keeps the physical size of the lattice fixed (so that the increasing number of lattice points makes the mesh size go to zero), the extrapolation length, measured in physical units, will also go to zero.

Note also that the distance in the $y$ directions was 511 rather than 512 because
we used
one row at $Y = 0$, in conjunction with the periodic boundary conditions,
to represent both horizontal boundaries
of the system on the lattice.  That is, $Y = 1$ and $Y = 511$ are the 
lowest and highest rows where we occupy sites in the system, where $Y$ 
represents the lattice coordinate in the vertical direction.

 We note
that these simulations were carried out before the
theoretical predictions were made.

\begin{figure}
\begin{center}
\includegraphics[width=5in]{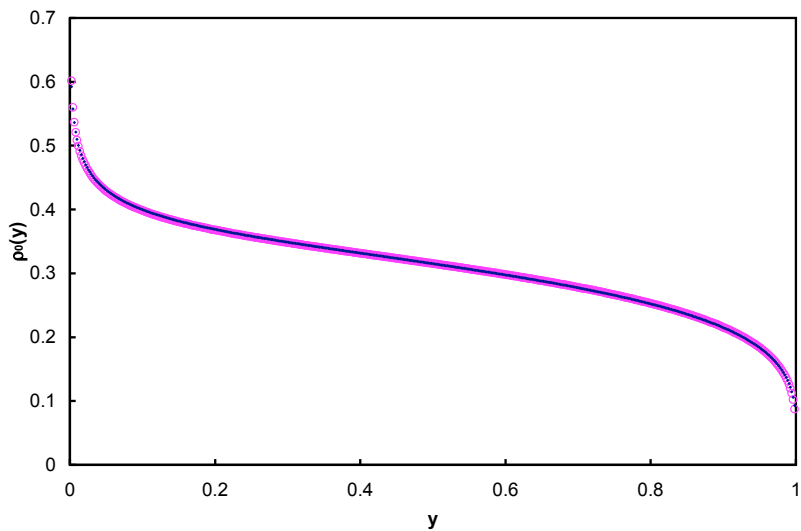}
\caption{ (Color online). 
Density of
 clusters touching lower boundary, $\rho_0(y)$, as a function
of $y$, both simulation (dots) and theory (open circles), equation (\ref{rho0}). }
\label{figrho0a0}
\end{center}
\end{figure}

We have also carried out simulations measuring the density of
clusters touching one edge, $\rho_0(y)$.  The results are shown in Fig.\ \ref{figrho0a0}.
We also plot the results of the theoretical prediction, equation (\ref{rho0}), on
the same plot, and
find agreement within 0.5\% without using an extrapolation length $a$.

\section{Percolation on a square and semi-infinite strip}

\begin{figure}
\begin{center}
\includegraphics[width=5in]{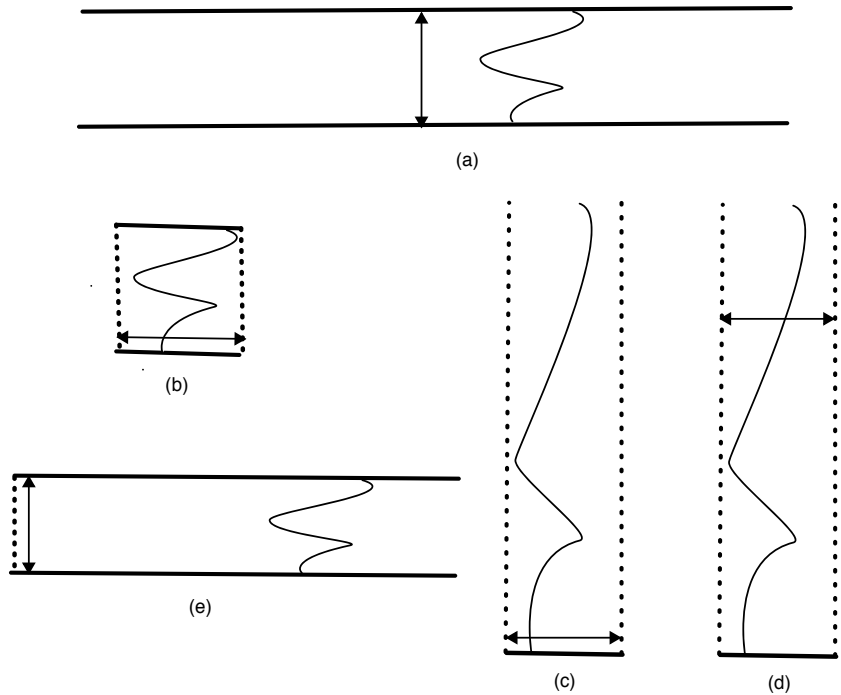}
\caption{Sketches of the cases considered.  Solid (dashed) boundary lines represent fixed (open) boundary conditions. Curved lines indicate crossing clusters; the density $\rho_b$ is evaluated along the lines with arrowheads. (a) is the infinite strip, cf. e.g. equation (\ref{rhoB}); (b) the square (Fig. \ref{figedgedensity}), (c) and (d) vertical half-infinite strips ((\ref{edgetheory2}) and (\ref{rhoBygg0}), respectively); (e) horizontal half-infinite strip (\ref{rhoBx0}).} \label{figx}
\end{center}
\end{figure}

In this section, we consider the density of crossing clusters on a square with open boundaries and also on a (long) rectangle.  We compare the numerical results with the predictions of conformal field theory for the density along an edge.  The various cases considered are illustrated in
Fig.\ \ref{figx}. 

Note that, as mentioned, the crossing is always in the vertical direction.  In a slight abuse of notation, we use  $\rho_b$ for the density of a clusters that touch both anchoring intervals in all cases.  The different situations may be distinguished by the arguments of $\rho_b$, e.g. $\rho_b(y)$ for the infinite strip, where there is no $x$ dependence, and $\rho_b(x,y=0)$ along the bottom of the semi-infinite strip as in (\ref{edgetheory2}) below.

\begin{figure}
\begin{center}
\includegraphics[width=5in]{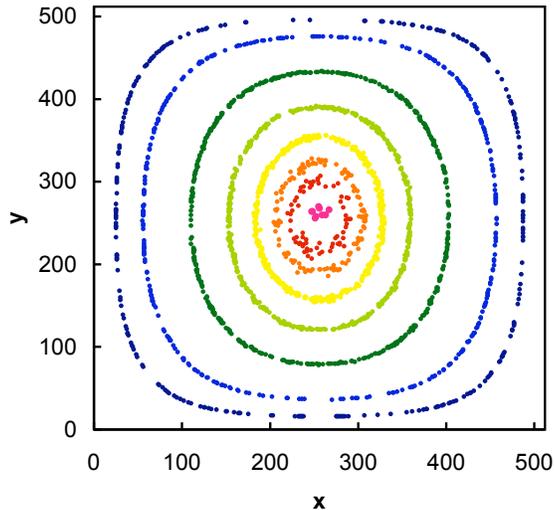}
\caption{ (Color online) Contours of constant densities 0.625, 0.75, 0.875, 15/16=0.9375, 31/32=0.96875, 63/64 = 0.984375, 127/128 = 0.9921875, and 1
(outside to center) of clusters touching both the top and bottom edges,
with open b.c.\ on the sides, for a system of $511 \times 511$ sites.} \label{squarecontour}
\end{center}
\end{figure}

 Our simulations of percolation densities
on an open square examined
clusters that cross in the vertical direction, with open boundary
conditions on the left- and right-hand sides.  We considered 
site percolation on a square lattice of $511 \times 511$, with 
2,000,000 samples generated.  The resulting contours
are shown in
Fig.\ \ref{squarecontour}.  As in the periodic case, the density
goes to zero at the upper and lower boundaries because, compared
to an infinite system, these boundaries intersect many possible
crossing paths, leading to large holes in the clusters near the 
boundaries.   As a consequence, relatively few points
on the boundary will be part of the crossing clusters, and in 
the limit that the mesh size goes to zero, their density evidently goes
to zero.
 The density
also goes to zero on the sides because of the open  conditions there.
Interestingly, the contour curves are almost symmetric in the horizontal
and vertical directions, indicating that the anchoring and open boundaries
have a similar effect on the density.

We have not carried out a field-theory calculation to find the density
inside the square or rectangular systems.
To do so  requires a six-point function
whose calculation would be unwieldy.

It is  however relatively easy to calculate the variation of the density along the bottom edge of the square, now normalized so that the density remains non-zero as
the mesh size goes to zero.  To do this, we consider the density of crossings from one of the anchoring intervals to a single point on the other interval, using the boundary spin operator.  Now crossing from the top edge to one point $x$ on the bottom edge (which is given by a three-point function, depending only on  $x$ and $\lambda$) automatically implies crossing from the top to bottom (which is given by a four-point function, depending only on  $\lambda$). Therefore the density at $x$ is proportional to the ratio of the former to the latter.  It follows generally that, up to a ($\lambda$-dependent) multiplicative constant, one has \cite{SimmonsKlebanZiff07}
\begin{equation} 
\rho(x) = \left( \frac{|z'(x)|}{1 - \lambda z(x)} \right)^{1/3} \ .
\label{edgetheory}
\end{equation}
Here  $z(w)$ maps the region of interest ($w$) onto the 1/2-plane ($z$), $x$ is the $w$-coordinate along the anchoring interval of interest, and $\lambda$ is the conformally invariant cross-ratio for the anchoring points.  For a rectangle, this depends on the aspect ratio $r =K(\sqrt{1-\lambda})/K(\sqrt{\lambda})$ \cite{Ziff95,Cardy92}.

\begin{figure}
\begin{center}
\includegraphics[width=5in]{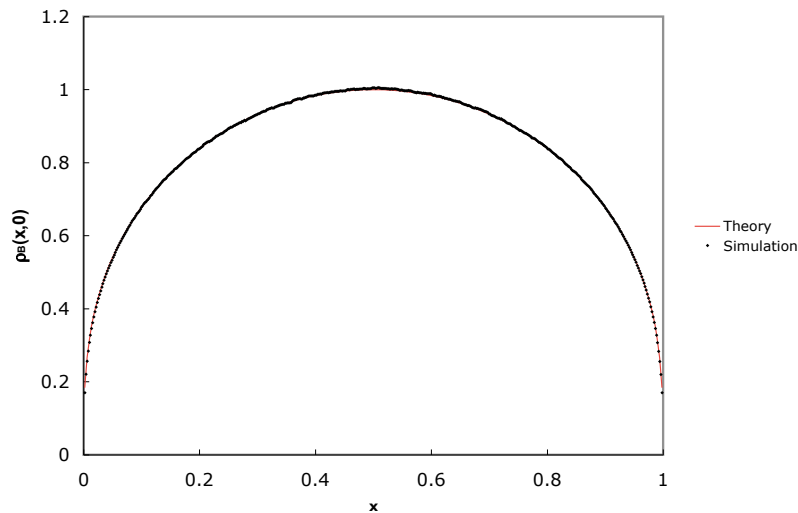}
\caption{ (Color online) Density of vertically crossing clusters $\rho_b(x,0)$ 
along the lower  boundary, in a square system with open boundaries
on the horizontal sides.  Red line: theory,
equation (\ref{edgetheory}), black dots: simulation
results.}
\label{figedgedensity}
\end{center}
\end{figure}

The mapping for the square is
\begin{equation}
z(w) = 1-\wp \left(\left(\frac{1}{2}+\frac{i}{2}\right) (i w+1) K(2);4,0\right) \ ,
\label{z}
\end{equation}
with $\wp(u; g_2, g_3)$ the Weierstrass  elliptic function and $K(z)$ 
the elliptic integral function.  This mapping takes a unit square $x, y \in (0,1)$ into the upper half plane. For the square $\lambda=1/2$, and we can take $x \in (0,1)$.
In Fig.\ \ref{figedgedensity} we compare
the measurement and theory. Clearly the agreement is 
excellent.

In the case of a half-infinite strip 
$x \in (0,1), y\in(0,\infty) $, 
the density of sites along the $x$-axis
belonging to clusters crossing vertically 
is found from (\ref{edgetheory}) using $z(w) = \sin^2(\pi w /2)$, and $\lambda = 0$.  This gives
\begin{equation}
\rho_b(x, y=0) = (\sin\pi x)^{1/3} \ .
\label{edgetheory2}
\end{equation}
Of course, for an infinite strip, the probability of crossing (in the long direction) is zero, 
so one must consider the limit of a large system, and calculate
the density {\it given that} crossing takes place, and take the limit
that the length of the rectangle goes to infinity.
It turns out that numerically, the density at the edge for the square,
equations (\ref{edgetheory},\ref{z}), differs only very slightly from 
that of the half-infinite strip, given by equation (\ref{edgetheory2}).  From the point of view of
the density along an anchoring edge, the square is not much different from the half-infinite strip.

For $y \gg 0$ in  the above half-infinite strip (or equivalently for any $y$ for a fully infinite strip in the vertical direction)
one can also find the density along the $x$-direction of the
vertically crossing clusters in closed form
\begin{equation}
\rho_b(x, y \gg 0) =(\sin \pi x)^{11/48} \ .
\label{rhoBygg0}
\end{equation}
This function may be found by transforming the density of clusters connecting two boundary points, derived in \cite{KlebanSimmonsZiff06}, into the infinite strip.  We then take the limit as the two anchoring points move infinitely far away in opposite directions, while normalizing the density so that it remains finite. (Related order-parameter profiles for the Ising case are studied in \cite{CFT97a}.)
Interestingly, a plot of  this density profile (written as a function
of $y$ rather than $x$) is very similar in appearance
to that of the vertically crossing clusters $\rho_b(y)$ given in equation (\ref{rhoB})
and plotted in Fig.\ \ref{figrhoBa0}.  When normalized so that they
have the same value at $y = 1/2$, the maximum difference between
the two is at $y=0$ and $y = 1$, where (\ref{rhoBygg0}) is only 1.5 \%
below (\ref{rhoB}).
This  small difference  indicates
 that open boundaries and anchoring boundaries
have similar effects on the density of the crossing percolation clusters, and is consistent with the near symmetry seen in the contours in Fig.\ \ref{squarecontour}.

We can also find the density along the left and right (open) edges.
For the case of a half-infinite strip 
rotated by $90^\circ$ with respect to the one above, ($y \in (0,1), x\in(0,\infty) $),
we find
\begin{equation}
\rho_b(x=0, y)=(\sin \pi y)^{-1/3}\left[\left(\cos \frac{\pi y}{2} \right)^{2/3}
+\left(\sin \frac{\pi y}{2} \right)^{2/3}-1\right] \;,
\label{rhoBx0}
\end{equation}
which is similar in form to $\rho_b(y)$ of equation (\ref{rhoB}) (which
in fact corresponds to $x \gg 0$ for the geometry considered here)
but with different exponents.   
This similarity arises because the derivations of (\ref{rhoBx0}) and (\ref{rhob}) are virtually identical, except that for (\ref{rhoBx0}) we leave a free interval between the two anchoring intervals (where  the boundary spin operator sits) which is mapped to the end of the half-infinite strip using sine functions.
Comparison with numerical data (not shown) for
the density at the short edge of a rectangle of
aspect ratio 8 to approximate the infinite strip
shows excellent agreement with (\ref{rhoBx0}).

\section{Further comments and conclusions}

If we consider densities raised to the sixth power
(compare Ref.\ \cite{KlebanSimmonsZiff06}), we find
a Pythagorean-like relation involving $\rho_e(y)^3$,
$\rho_0(y)^3$, and $\rho_1(y)^3$ (which we present without interpretation),
\begin{equation}\label{Pyth}
\rho_0(y)^6 + \rho_1(y)^6 = \rho_e(y)^6 \;.
\end{equation} 
There seems to be no simple relation involving $\rho_b(y)$
other than its basic definition given in equation  
(\ref{rhob}).   For the corresponding quantities at the edge
of the strip as in (\ref{rhoBx0}), the power in equation 
(\ref{Pyth}) is 3 rather than 6.

Although  the overall  normalization of  a density such as
$ \rho_b(y)$ (for the infinite strip, see (\ref{rhoB}))  is arbitrary,
we can fix its value by requiring that
\begin{equation}
\int_0^1 \rho_b(y) dy = 1 \;.
\label{integralB}
\end{equation}
That is, we define 
 $\rho_b(y) = B \sin(\pi y)^{-5/48}(\cos(\pi y/2)^{1/3}+\sin(\pi y/2)^{1/3}-1)$
where $B$ is a constant.  Then equation (\ref{integralB}) yields
\begin{eqnarray}
B &=& 
   \left[ -\frac{\Gamma \left(\frac{43}{96}\right)}{\sqrt{\pi }
   \Gamma \left(\frac{91}{96}\right)}+\frac{32 \Gamma
   \left(\frac{59}{96}\right) \Gamma
   \left(\frac{43}{48}\right)}{\sqrt{\pi } \Gamma
   \left(\frac{1}{16}\right) \Gamma
   \left(\frac{91}{96}\right)}\right] ^{-1}\\
   &=& 1.46408902\ldots \;.
 \end{eqnarray}
 Another choice of $B$ is to make $\rho_b(1/2) = 1$, which yields 
$B = (2^{5/6}-1)^{-1} = 1.27910371\ldots$

In many problems   of percolation density
profiles,  the density goes to 
infinity at a boundary point, such as occurs for
$\rho_0(y)$ when $y \to 0$.
 Interestingly, in the case considered here of clusters touching both boundaries,
$\rho_b(y)$, the density goes to zero at  those boundaries
and remains finite everywhere.

To highlight the difference between densities
of all clusters touching a boundary vs.\ the densities of crossing clusters
touching one boundary,
we consider the limit that the strip width
becomes infinite, so that the system becomes a half-plane.
Because we have written our results for a strip of fixed (unit)
width, this density is given by 
the behavior of $\rho$ for small $y$.
The density of {\it all} clusters touching the $x$-axis is
thus found from equation (\ref{rho0}) (or \cite{BE85,ResStraley00}) to be 
\begin{equation}
\rho_0 \sim y^{-5/48} \ ,
\label{rho0strip}
\end{equation}
where we have left off the coefficient because the normalization
is arbitrary.  This density diverges at $y = 0$ because it is much
more likely to find sites belonging to these clusters near the
$x$-axis.  On the other hand, the density of {\it crossing} clusters touching
the $x$-axis is found from  equation (\ref{rhoB}) in the limit $y \to 0$, 
\begin{equation}
\rho_b \sim y^{11/48} \ .
\label{rhoBinfinite}
\end{equation}
In this case, the density increases as $y$ increases, in contrast to (\ref{rho0strip}),
and 
goes to zero at the  anchoring boundary, for the reason mentioned above.

Behavior identical  to (\ref{rhoBinfinite})
also follows from the small-$x$ expansion
of $\rho_b(x,y \gg 0)$ given by
(\ref{rhoBygg0}), which represents the behavior of
the density of the vertically spanning clusters at
the open boundary.  Thus, near the boundaries
(but away from the corners), the
open and anchoring boundary conditions have identical effects
on the density of the crossing clusters.

 In conclusion, we have studied the density of vertically
percolating clusters in a square system, as well as for
 half-infinite and infinite strips extending
in either the horizontal or vertical direction. The various cases considered are illustrated in Fig.\ \ref{figx}. 

For the infinite strip, Fig.\ \ref{figx}(a), the density for crossing clusters is given by  (\ref{rhoBx0}) and compared with numerical simulations in  Figs.\ \ref{figrhoBa0} and \ref{figratioboth}.
For the square, Fig.\ \ref{figx}(b),
we find theoretical results for the anchoring
edge densities (see (\ref{edgetheory}) and (\ref{z}) and Fig.\ \ref{figedgedensity}), as well as numerical results for the density in the interior  (Fig.\ \ref{squarecontour}), which exhibit an interesting near symmetry.
For the half-infinite strip in the horizontal direction, Fig.\ \ref{figx}(e),
the density $\rho_b(x=0,y)$ at the left open boundary is given by (\ref{rhoBx0}),
and at $x \gg 0$ (or equivalently, for an infinite strip), the
density is given by (\ref{rhoB}).  For a half-infinite strip in the vertical direction, the density along the
lower anchoring boundary (Fig.\ \ref{figx}(c)) is given by (\ref{edgetheory2})
while for $y \gg 0$ (Fig.\ \ref{figx}(d)) the density is given by (\ref{rhoBygg0}).
For the half-infinite systems, the densities near a
wall are given by the same power-law (\ref{rhoBinfinite})
regardless of whether it anchors the crossing clusters or is open.
Note that all of our theoretical predictions were confirmed by 
computer simulation.  

 For the future, it would be interesting to 
study analogous properties for Fortuin-Kasteleyn (FK)
clusters of the critical Ising and Potts models.

\section{Acknowledgments}
This work was supported in part by the National Science Foundation under grants numbers  DMS-0553487 (RMZ) and DMR-0536927 (PK).

\end{document}